\documentclass[a4paper]{jpconf}
\usepackage{amsmath}
\usepackage{amssymb}
\usepackage{epsfig}
\usepackage{color}
\usepackage{graphicx}

\begin{document}

\title{Neutrinos from SN 1987a \\ A Puzzle Revisited}

\author{Gerd Schatz}

\address{Universit\"at Heidelberg \textit{and} Karlsruhe Institute of Technology}

\ead{bgschatz@t-online.de}

\begin{abstract}
The smallest of the four detectors which claim to have observed neutrinos from SN 1987a registered the events more than 4 h earlier than the other three ones. This claim is not usually accepted because it is difficult to understand that the other (and larger) detectors did not register any events at the same time. It is shown that microlensing of the neutrinos by a star in-between the supernova (SN) and Earth can enhance the neutrino intensity at the position of one detector by more than an order of magnitude with respect to the other detectors. Such a configuration is improbable but not impossible. Essential for this enhancement is the small source diameter, of order 100 km. So if two bursts of neutrinos were emitted by SN 1987a at a separation of about 4 h it could be explained easily that the smallest detector observed the first burst while the other ones missed it and vice versa.
\end{abstract}

\section{Introduction}

Supernovae of type II, also called core collaps SN, convert of order one solar mass of protons into neutrons. They therefore produce a neutrino burst of several tens of seconds duration. This was observed once so far when the famous SN 1987a flared up in the Large Magellanic Cloud on February 23, 1987. One of the most conspicuous features of these observations is the fact that the smallest detector registered 5 events more than 4 h earlier than the other three. The following table gives some characteristics of the four detectors and summarizes the observational results (data from \cite{imb:87} \cite{kam:87} \cite{bak:88} \cite{lsd:87} \cite{olga}). The claim of the LSD collaboration that the 5 events observed originate from SN1987a was soon called \textit{unbelievable} \cite{aww} since it appeared unexplainable that the other three (and larger) detectors did not see any events at the same time. It is the purpose of this contribution to demonstrate that a mechanism exists which could explain this discrepancy in a natural way.
it is not our aim to discuss the contested question of the SN 1987a neutrinos in any comprehensive way but only to draw attention to a special aspect which apparently has been overlooked in the past but is of relevance for the problem.

\begin{center}
\begin{table}[h]
\centering
\begin{tabular}{|l|c|c|c|c|}
\hline
 & & & & \\
\textbf{
Detector} & \textbf{Size} & \textbf{Threshold} & \textbf{Number} & \textbf{Time of first} \\
 & \textbf{[t]} & \textbf{[MeV]} & \textbf{of events} & \textbf{event [U.T.]} \\
  & & & &\\
\hline
   & & & &\\
\textbf{ IMB} & 5000 & 15 & $~~8$ & $ 7^h~35^m~41^s$\\
  & & & & \\
\textbf{ Kamiokande} & 2140 & $~~~~7.5$ & 12 & $\approx~7^h~35^m $ \\
  & & & & \\
\textbf{ Baksan} & $~~200$ & 10 & $~~5$ & $ 7^h~36^m~12^s $\\
  & & & & \\
\textbf{ LSD} & $~~~290$ & $~~5$ & $~~5$ & \textcolor{red}{$ 2^h~53^m 11^s$}  \\
  & & & & \\
  & & & & \\
\hline
\end{tabular}
\end{table}
\end{center}

\section{Microlensing of neutrinos}

Gravitational deflection of light by a star can lead to a transient increase of the observed intensity of a background object. This effect is called microlensing and has been observed frequently in the optical range. The  signal has a well defined time dependence and is strictly achromatic. Fig. 1 illustrates the situation schematically. Since all extremely relativistic particles move in gravitational fields on orbits identical to those of photons the same should hold for neutrinos from SNe.

\begin{figure}
\begin{center}
{\includegraphics[scale =1.1]{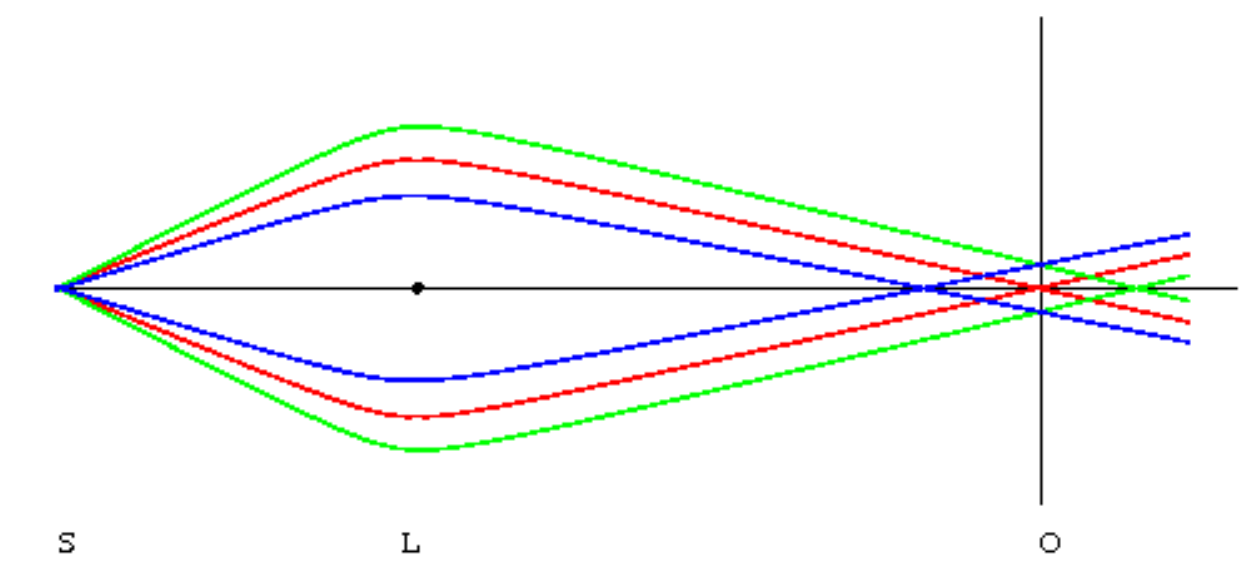}}
\end{center}
\caption{Schematic representation of microlensing with source S, lens L and observer O.}
\end{figure}

The effect of microlensing is an increase of the (light or neutrino) intensity in the plane perpendicular to the axis defined by source and lens at the distance of the observer. The time dependence results from the movement of the observer across the radiation field (stationary in the case of light). Since neutrinos and photons have identical orbits we can use all formulae derived by astronomers also for neutrinos. The radial dependence of the intensity in this plane can be calculated analytically for point sources \cite{Schn:92} \cite{Pacz:96}. The result is shown in fig. 2. It tends to a constant value, undisturbed by the lens, at large distances from the axis and increases inversely proportionate to the distance $r$ from the axis near the center.

\begin{figure}
\begin{center}
{\includegraphics[scale =1.]{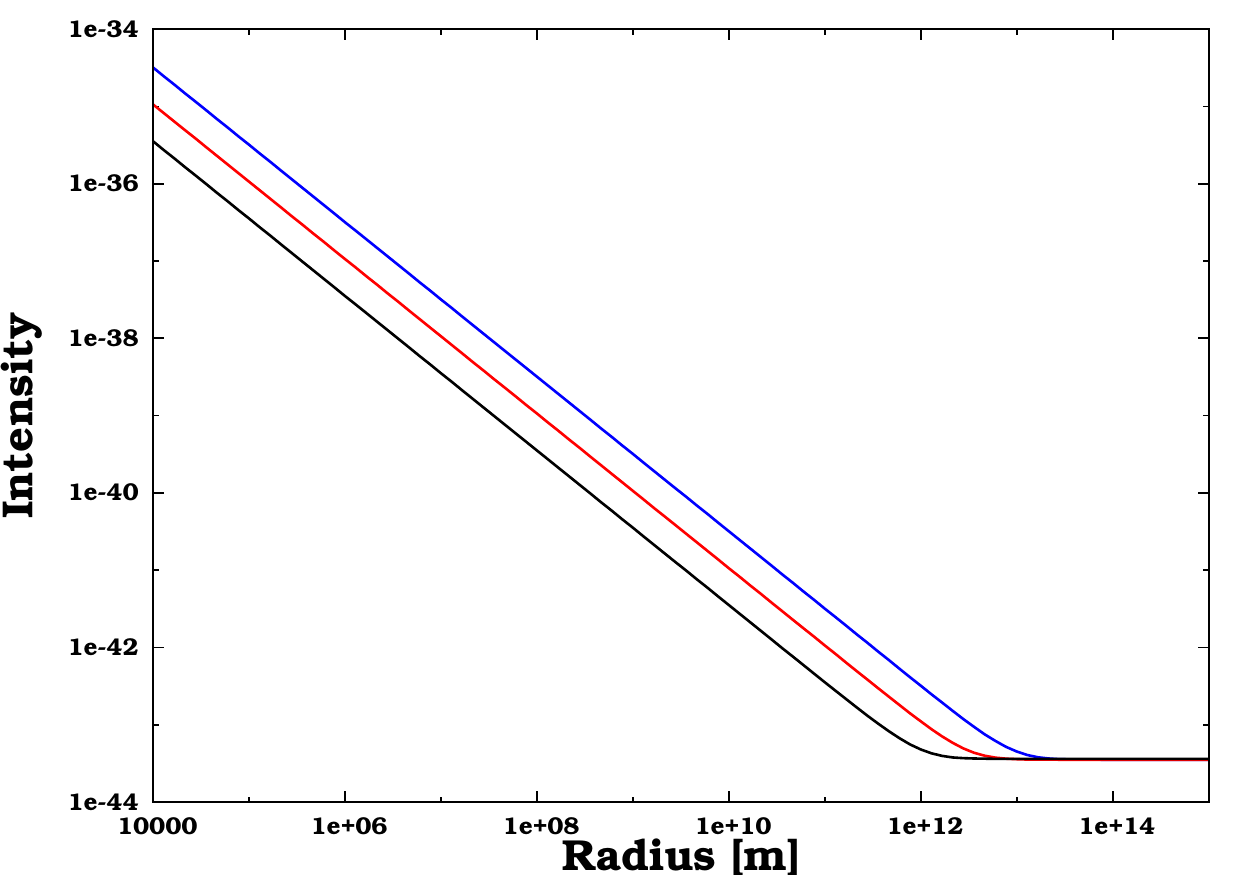}}
\end{center}
\caption{Dependence of intensity on radius of the observer for a point source. Different colours represent different lens positions at 10 \%, 50 \% and 90 \% of the source-to-observer distance.}
\end{figure}

The ratio of the intensity to its undisturbed value is called amplification and given by


\begin{equation*}
\begin{split}
A(r)~=~\frac{I(r)}{I(\infty)}~&=~\frac{1}{2~a~r}~
\frac{~R_E^2~+~2~a^2~r^2}{\sqrt{~R_E^2~+~a^2~r^2~}}
\end{split}
\end{equation*}

\noindent for point sources. Here $r$ is the distance of the observer from the axis, $R_E=\sqrt{2 R_S D_{SL} D_{LO}/D_{SO}}$ is called the Einstein radius, $R_S$ is the Schwarzschild radius of the lens, the $D$s are the distances between source, lens and observer as indicated by the indices, and $a=D_{SL}/2 D_{SO}$. For large radii $r \gg R_E/a$ the amplification tends to 1 and for small ones it is inversely proportionate to $r$. \\

The singularity on the axis disappears when this formula is folded with the finite size of the source. The result is an effective truncation of intensity and amplification with a maximum on the axis of $A(0)=R_ED_{SO}/(r_sD_{LO})$. Here $r_s$ is the source radius. For $r \gtrsim r_s D_{LO}/D_{SL}$ the point size expression holds to  a good approximation. The intensities after folding are shown in fig. 3 for sources of the sizes of a SN and the Sun. They are approximately constant near the axis and at large distances and decrease inversely proportionate to $r$ in-between.

\begin{figure}
\begin{center}
{\includegraphics[scale =1.]{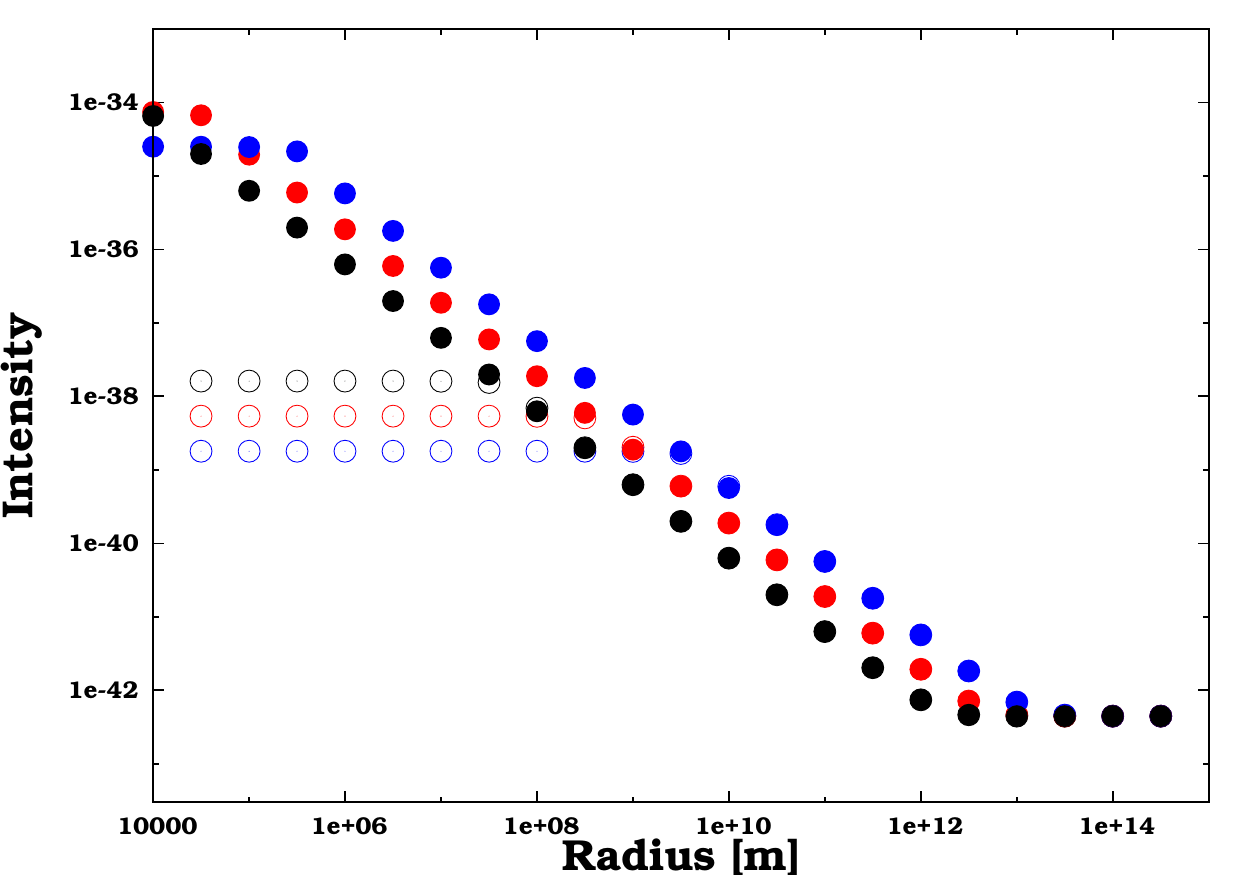}}
\end{center}
\caption{Neutrino intensities in the observer plane for sources of the size of a SN (full symbols) and the Sun (open symbols). The colours have the same meaning as in fig. 3.}
\end{figure}

\section{Application to SN 1987a}

At the moment of observation of SN 1987a the separations between the LSD detector and the other ones, projected on a plane perpendicular to the line connecting the SN with the Earth, amounted to approximately 2800 [km] (Baksan), 6100 [km] (IMB,) and 8700 [km] (Kamiokande). If the LSD detector is situated in the central plateau of the intensity distribution and another detector D in the decreasing part the ratio of amplifications is given by $2r_DD_{LO}/(r_sD_{SL})$ where $r_D$ is the distance of the second detector from the axis. If both detectors are placed in the sloping part the ratio is simply $(r_D+r_{LSD})/r_D$. In either case it is easy to find configurations which enhance the LSD signal with respect to the other ones by an order of magnitude or more. This would be  sufficient to explain that the LSD detector saw the first neutrino burst while the other ones failed to register any events not the least in view of the small event numbers involved and the correspondingly large statistical errors. Since the observer is expected to move with respect to source and lens there is a high probability that the LSD detector has moved to outside the cone of enhancement 4 h later. This would result in a high probability for the LSD detector of missing the second burst. \\

It is worthwhile mentioning that the described mechanism does not require exceptionally large or heavy stars as gravitational lenses. Even a planet of the size of Jupiter would do if placed at more than $\approx 10^3$ [A.U.] from the source. One should also be aware that the probability of the precurser star having companions ranging from planets to a neutron star is very high. \\

\section{Discussion}

There is no doubt that the probability of such a configuration occurring at the moment of SN explosion is very small. We refrain from trying to estimate this probability quantitatively since this would require assumptions on the spatial and mass distributions of possible lenses which are to a considerable extent arbitrary. It would in any case be much smaller than the probability of the LSD events being a statistical fluctuation which is given by $4 \cdot 10^{-4}$ \cite{lsd:87}. But it is also clear that the required configurations are not impossible. \textbf{Hence one has to conclude that the argument against the LSD claim is not as compelling as has hitherto been assumed.} In view of the small probability of the required configurations occurring it is of course still up to the reader whether to consider the claim \textit{unbelievable} \cite{aww} or not. \\

An unresolved problem which remains is the cause for two neutrino bursts originating from the SN. Hillebrandt et al. \cite{hil} proposed a mechanism which relies on the fact that all stars rotate to some extent. This will cause an acceleration of the rotational movement (pirouette effect) with a corresponding increase of centrifugal forces which might halt the collaps for a certain period of time until some angular momentum has been lost by interaction with the environment. This might then result in a two-step collaps. An alternative mechanism invoking fragmentation of the collapsing star has been discussed more recently \cite{olga}. These explanations appear to have lost appeal, though, because the SN precursor apparently loses a considerable fraction of its angular momentum during the giant phase immediately preceding the collaps \cite{jan}. It is not unfair, though, to add that the theory of type II SNe is hardly in a definite final state so revision of this argument does not appear impossible. \\

\noindent \textit{Acknowledgements}

\noindent I thank M. Bartelmann (Heidelberg) for patient instructions concerning general relativity, T. Huege (Karlsruhe) for equally patient assistence during my fight with  \textit{gnuplot}, and D. Huber (Karlsruhe) for his indispensible help in producing the poster displayed at the conference. \\

\end{document}